\documentclass[cameraready]{Interspeech}

\title{The Last Mile of Deepfake Speech Detection:\\ An Industry–Academia Experience Report}

\author[affiliation={1}, orcid=0000-0002-4717-1910, correspondingauthor]{Anton}{Firc}
\author[affiliation={1}, orcid=0000-0002-9009-2193]{Kamil}{Malinka}
\author[affiliation={1}, orcid=0009-0000-5722-0571]{Vojtěch}{Staněk}
\author[affiliation={2}]{Miroslav}{Hlaváček}
\author[affiliation={2}]{Marek}{Bartoň}

\address{
    $^1$ Security@FIT, Brno University of Technology, Czech Republic \\
    $^2$ Phonexia, Brno, Czech Republic
}

\email{ifirc@fit.vut.cz, malinka@fit.vut.cz, istanek@fit.vut.cz}


\usepackage{comment}

\usepackage{xspace}
\usepackage{xurl}

\makeatletter
\@ifundefined{ifcameraready}{%
  \newif\ifcameraready\camerareadyfalse
}{}
\makeatother

\ifcameraready
  \newcommand{\company}{Phonexia\xspace}
  \newcommand{\institution}{Brno University of Technology\xspace}
  \newcommand{\useragency}{the Czech Police\xspace}
\else
  \newcommand{\company}{[Company]\xspace}
  \newcommand{\institution}{[University]\xspace}
  \newcommand{\useragency}{[User organisation]\xspace}
\fi

\begin{document}

\maketitle

\begin{abstract}
Synthetic speech detection benchmarks now report sub-1\% error rates on some in-domain evaluations, yet performance degrades under unseen attacks, channel mismatch, and distribution shift. Based on a three-year effort with \company, a commercial speaker-recognition vendor, we report barriers encountered while building and deploying a detector. Many public benchmarks are not licensed for commercial model development. Real inputs are not four-second clean clips but long, codec-degraded, sometimes partially synthetic recordings. And when a calibrated system returns a log-likelihood ratio of 2.5, no one can tell the customer what it means for their decision. Rather than proposing a new model, we connect these barriers to concrete research and coordination proposals: shared standards for commercially usable datasets, realistic deployment benchmarks, and scores that non-experts can act on. These observations come from one project and should be tested in other settings.
\end{abstract}

\section{Introduction and Position}
\label{sec:intro}

By standard benchmark measures, deepfake speech detection appears close to saturation. Leading systems report error rates well under 1\% on in-domain benchmark data, and each development cycle further narrows the gap~\cite{wang2024asvspoof5}. Those same systems, however, degrade sharply on data drawn from outside their training distribution~\cite{muller2022does}. Strong in-domain discrimination, therefore, does not establish robustness or readiness for deployment. Synthetic speech has already been shown to threaten speaker-verification systems and consumer voice assistants~\cite{firc2022dawn,malinka2024voiceassistants}. This paper reports the technical, operational, and organizational barriers we encountered while turning a detector into a product.

We learned this by trying it ourselves. Over the course of a three-year project to bring a commercial detector to market, the recurring difficulties extended beyond the model architecture. Several of the most persistent problems concerned the systems and processes around the model:
\begin{itemize}[topsep=0pt, itemsep=0pt]
    \item training data we were legally permitted to use,
    \item evaluation that says something about the conditions a customer actually meets rather than a leaderboard,
    \item calibration for deployments without labeled data,
    \item scores a non-expert can turn into a decision,
    \item and integration into systems we never see.
\end{itemize}

Our evidence comes from one project involving one university, one vendor, and one user organization. We therefore report these observations as project-specific evidence and use them to motivate hypotheses and proposals for similar deployments, rather than as findings that apply to every vendor or user organization.

Existing work approaches the problem from several angles. Most of the literature advances detection methods, surveyed in~\cite{deepfkeSurvey} and driven by the ASVspoof challenge series~\cite{yamagishi2021asvspoof}. A second strand documents the generalization gap, showing that detectors that excel on a benchmark falter on unseen attacks and on genuinely in-the-wild material~\cite{mlitw2026}, which in turn has motivated larger and more diverse corpora spanning many languages and synthesizers~\cite{muller2024mlaad, jung2025spoofceleb}. A third examines the ethical and licensing limits of the data on which the field depends~\cite {stanek2026datasets}. Closest to our setting, recent work analyzes commercial deepfake detectors on real-world cases~\cite{muneer2025commercial}. That work evaluates finished systems from the outside, whereas we report the development and deployment process from the inside, including the choices, constraints, and unsuccessful approaches encountered by a vendor and its academic partners. This paper is therefore an experience report rather than a new-model study. Its contribution is a structured account that connects deployment observations to open research questions and proposed coordination actions. To our knowledge, this productionization process has not been documented systematically before.

\textbf{\textit{Our position}} is that improvements in in-domain benchmark accuracy alone are insufficient for dependable deployment. Our project exposed unresolved technical challenges in generalization, calibration, and robustness, as well as operational and collective constraints in data access, evaluation conventions, and score communication. Because no single organization controls all the shared resources and conventions involved, progress requires both further research and community coordination. This paper distills the obstacles we encountered into open problems and proposes actions, and makes the following contributions:

\begin{itemize}
    \item \textbf{An experience report} from an active commercial productionization effort, jointly authored by \institution and \company, documenting the barriers encountered while bringing this detector to market.
    \item \textbf{An evidence-linked roadmap of open challenges} that maps project observations to unresolved questions in data selection, evaluation, generalization, calibration, score communication, integration, governance, and customer acceptance.
    \item \textbf{A call to action} translating the roadmap into concrete, mechanism-level proposals the community can act on.
\end{itemize}

\section{Industry-Academia collaboration}





The work reported here arose from a three-year project involving \institution, \company, and \useragency as the user organization. The university and the company each had a clear reason to pursue it, though not the same one. \company's customers had begun asking whether its speaker-recognition products could also detect deepfakes; they could not, and adding that capability was a natural extension of the company's existing product portfolio. For the university, audio deepfake detection was a new and largely open research problem. Because \company is a university spinoff and the two have kept close ties, establishing a joint project was straightforward.


Funding was provided by the Czech Ministry of the Interior through the SECTECH security research program for the project \emph{Tools to Combat Voice DeepFakes} (VB02000060)~\cite{mvcr_program,project_fit}. The Czech Police served as the end user, and the concrete deliverable was a set of tools they could apply in their work. This framing kept the project focused on operational deployment rather than benchmark performance alone.

The division of labor reflected each partner's strengths. The university contributed the research: model design, current methods, and an informed view of where the field actually stood. \company served as the implementation partner, turning that research into a system that ran end-to-end and could be delivered to \useragency, which supplied the operational context and the requirements that any tool intended for them had to meet. None of the three partners could have produced the result alone, and each gained more from the collaboration than it contributed in isolation. For a problem that sits precisely on the boundary between open research and operational deployment, dividing the work along these lines proved effective: each partner covered the part the others could not.

\section{Experience Report}
\label{sec:experience}

This section reports on the issues that dominated productionization in this project. The account is retrospective and project-specific; it does not estimate how often these issues occur across vendors or deployment settings. For each theme, we describe what we encountered, what made it hard, what we tried, and, where disclosure permits, concrete results from \company.

\subsection{The Data Problem}
\label{sec:data}

The literature has repeatedly documented that competitive detectors trained on the canonical benchmarks degrade catastrophically when evaluated on synthesized audio they have not seen during training. M\"uller et al.~\cite{muller2022does} reported equal-error-rate degradations of up to 1000\% between in-domain ASVspoof~2019 evaluation and a curated in-the-wild celebrity deepfake dataset. Subsequent benchmarks confirm the pattern across ASVspoof~2021~\cite{yamagishi2021asvspoof}, ASVspoof~5~\cite{wang2024asvspoof5}, and multilingual in-the-wild conditions~\cite{mlitw2026}. The current community consensus is that speech deepfake detection is fundamentally an out-of-distribution generalization problem: detectors trained on a finite set of synthesizers must generalize to ones they have never seen at deployment~\cite{ibcaan2025}.

The implication for a commercial detector is direct: training and evaluation data must track the leading edge of speech synthesis, because customers will benchmark against whichever synthesizer has the most public visibility at the moment of evaluation. Diffusion-based speech illustrates this moving target: its detectability varies across detector architectures, even when its overall detectability is comparable to that of non-diffusion speech~\cite{firc2024diffuse}. A particularly visible example in 2026 is ElevenLabs.

\subsubsection{Licensing and the gray zone}
\label{sec:data-licensing}

Commercial access to current TTS outputs is constrained by provider policies. For example, ElevenLabs' publicly posted policy prohibits using its services or outputs as input to machine-learning models and prohibits including them in datasets used to train, test, or improve such systems~\cite{elevenlabs_policy}. The scope of these restrictions differs across providers~\cite{openai_terms,playht_terms}. Their legal effect may also depend on the applicable contract, separate authorization, jurisdiction, and intended use. In our project, this uncertainty meant that outputs from relevant commercial synthesizers could not be assumed to be available for detector training or evaluation.

This creates a practical mismatch. Detector developers need data from the synthesis systems that appear in their threat model, but the publicly available terms may restrict the collection or use of those outputs. The enforceability of terms of service and end-user license agreements, and their interaction with intellectual-property law, varies across jurisdictions~\cite{oecd2025scraping}. We do not assess the legality of specific data-acquisition practices. Our narrower observation is that developers lack a transparent and standardized route for obtaining commercially usable, deployment-relevant data.

Several detection vendors advertise coverage of outputs from leading commercial synthesizers, including ElevenLabs~\cite{corsound2026guide,aurigin2026}. These public statements do not reveal whether the synthesizer outputs were used for training or testing, how the relevant data were obtained, or whether separate permissions were granted. We therefore draw no conclusion about any vendor's compliance. The relevant problem for this report is the lack of transparent data provenance: product descriptions and deployed models provide little information about the data used to develop them.

We interpret this as a data-access and transparency problem rather than evidence of wrongdoing. Commercially usable and representative data remain difficult to obtain at the required scale.\footnote{A detailed survey of the licensing status of publicly available audio deepfake corpora is the subject of separate work~\cite{stanek2026datasets}.} When deployment-relevant data are unavailable, developers may have to train and evaluate on data that poorly match the intended use. This mismatch can weaken real-world performance independently of the legal status of any specific dataset.

\subsubsection{Data quality and drift}
\label{sec:data-quality}

Bona fide data also vary across sources and operating conditions. In our experience, a detector can learn characteristics of one bona fide distribution and perform poorly when evaluated on another. This mismatch also affects generalization testing: Staněk et al.~\cite{stanek2026datasets} show that many datasets derive from the same underlying corpora, which can introduce leakage and bias during training and evaluation.

A further complication is the versioning of the synthesis tools themselves. Treating a synthesizer such as ElevenLabs as a single, fixed target is misleading: its versions (v1, v2, v3, and the updates in between) differ in quality and in the artifacts they leave behind, so a detector trained on one version offers no guarantee of catching successive versions.

Similar instability also affects the bona fide data. Audio sourced from platforms such as YouTube is subject to continual changes in compression and codecs, and a corpus collected at one moment slowly drifts away from what the platform actually delivers today. We encountered this directly when we set out to evaluate the detector on YouTube data: the only usable material available to us was roughly eight years old and bore little resemblance to current YouTube audio, so the experiment told us essentially nothing. Both observations lead to the same conclusion: training and evaluation data have a limited shelf life, and tracking the leading edge is a moving target on two axes at once, i.e., the synthesizer and the channel.

These continuous updates raise a question: given what data is available and how quickly it dates, how do we decide which of it we actually need? The current practice is simply to use everything we have, which is the safe default but not an efficient one. It results in accumulating ever-larger amounts of data, requiring considerable computing power on material that may not provide much. A more principled approach would assess which data genuinely improve the detector, identify where coverage is still missing, and discard or undersample redundant data. While there is some initial research regarding data drift~\cite{wang2025datadrift} and active data selection~\cite{wang2022dataselection, furuhari2024dataselection}, we treat this as an open problem, and return to it in our call to action (\autoref{sec:cta}).

Finally, channel coverage, especially the inclusion of telephony-band audio in training, proved to be one of the most beneficial data choices we made; we treat its effect, which is most visible as a robustness issue, in \autoref{sec:eval-robustness}.

\subsection{Architecture and Model Selection}
\label{sec:architecture}




Selecting an architecture proved less settled than the literature suggests. The field offers several competing backbone families: self-supervised (SSL) speech representations~\cite{tak2022wav2vec, wang2022ssl}, raw-waveform end-to-end models~\cite{tak2021rawnet2, jung2022aasist}, and spectrogram-based convolutional networks~\cite{lavrentyeva2019stc, he2016resnet}. The difficulty is that each tends to perform best on a different benchmark, and in our experience, none dominates across the conditions a customer actually encounters. A model that tops one evaluation can fall behind on another, and in-domain equal-error rate is a poor predictor of cross-domain robustness. Architecture selection is therefore a question of which trade-off to accept, not which model is best outright~\cite{rohdin2024but,stanek2026fusion,firc2026spaarsist}.

We approached it in stages. We began with several open-source architectures, which gave us a baseline and a sense of where the easy gains and the hard limits lay. The research group at \institution{} then contributed two successive generations of models from its own work, along with ongoing support in adapting them. \company{} integrated these into its training and inference pipelines, retrained them on its own data, and optimized them for the target deployment.

Across variants, we settled on pretrained SSL front-ends with attentive pooling~\cite{peng2023mhfa}. SSL representations are learned from large amounts of unlabeled speech, which gives the detector a better starting point for generalizing to audio and synthesizers it has not seen during training, the property that mattered most to us given the out-of-distribution nature of the problem (\autoref{sec:data}). At inference time, each recording is processed chunk by chunk, and the final score is the maximum across chunks. This lets the detector flag partial spoofs, where only a segment of an otherwise genuine recording is synthetic, a case our forensic users care about in particular.

\subsection{Evaluation in Deployment Conditions}
\label{sec:evaluation}

Evaluating a detector for deployment was one of the hardest parts of the project, and the part where the distance between benchmark practice and operational reality was the widest. Four difficulties recurred:
\begin{enumerate}
    \item building an evaluation dataset that resembles what a customer will actually encounter,
    \item choosing metrics that mean something at the operating point a customer runs,
    \item establishing whether the system genuinely generalizes to unseen attacks, and
    \item understanding how sensitive the system is to channel and processing conditions.
\end{enumerate}

\subsubsection{Building evaluation data}
\label{sec:eval-data}

There is no settled answer to how much test data is enough, or how it should be composed, and we went through several iterations of evaluation sets. The goal is easy to state: \textit{as many and as diverse samples as possible}, but it explodes quickly: every additional synthesis tool, speaker, language, or codec enlarges the space to be covered. We therefore deliberately chose breadth over depth. We use a relatively small number of speakers and source corpora (30 speakers), but sweep several configurations of each synthesis tool, including custom/finetuned configurations, to ensure the detector is not tuned only to a tool's default settings.

On top of this, we add a basic set of augmentations, primarily telephony and background noise, along with simple manipulations that occur naturally or that an attacker could apply intentionally. This favors catching a wide range of attack types over detecting any single one perfectly. The data is suitable for a reasonable deployment, but it is an explicit trade-off rather than a principled approach, and we do not have a good answer for what evaluation breadth or depth is sufficient.

\subsubsection{Metrics that do not transfer}
\label{sec:eval-metrics}

Prior comparative work evaluated 40 detectors under a shared framework and found substantial variation in robustness across methods and input modifications~\cite{firc2025evaluation}. Deployment adds a further problem: the metrics the community optimizes do not correspond directly to the operating point a customer uses. Equal-error rate (EER), the headline number on every benchmark, says nothing about the operating point a customer actually runs. The detection cost function (DCF)~\cite{brummer2006}, the other standard benchmark metric, does pin down an operating point by weighting misses and false alarms against a prior, but that prior and those costs are fixed by the benchmark, not the customer, so even a strong minDCF says little about behavior at the threshold a deployment actually uses. In both cases, the false-alarm cost that dominates real deployments stays effectively invisible on a leaderboard.

We tried two things. First, because not every sample matters equally to a customer or to us, we weight samples by how important or interesting we judge them to be and report weighted metrics. This is openly ad hoc, an in-house heuristic that approximates what our customers and we perceive as important, rather than anything principled. Importantly, this is a deficiency that many others in the field also face.

Second, and more troubling, aggregate numbers hide a threshold problem. Each attack's EER is measured at whatever threshold is best for that attack, so a set of attacks can each report a low EER, say under 5\%, while each achieves it at a different threshold. But a deployed detector runs at a single fixed threshold, not a different one for each attack. Forced onto a single threshold, the combined system performs noticeably worse than the per-attack figures suggest, because no one threshold is right for all of them. The same gap appears across datasets: an EER reported separately on each dataset, with no threshold information, gives no way to tell whether the numbers are even roughly comparable, so a single EER per dataset is close to meaningless for someone deciding how to deploy. This is not a new concern in biometrics: ISO/IEC 19795~\cite{iso19795} already prescribes reporting error rates across a range of operating points rather than at one threshold. Speech deepfake detection has largely not adopted this, and closing the gap between single-number leaderboard reporting and operating-point-aware evaluation is something the community could do immediately.

\subsubsection{Testing generalization to unseen attacks}
\label{sec:eval-generalization}
A deeper problem is that we cannot be certain we are testing generalization at all. A real deployment must cover a far wider range of attacks than any single benchmark provides, and the obvious response is to combine several datasets. But each dataset brings its own labeling scheme, and there is no reliable way to tell whether, say, attack \texttt{A19} in one corpus and \texttt{myTTS\,+\,Malacopula}~\cite{todisco2024malacopula} in another are the same underlying system or different ones. Pooling the data and drawing random splits does not guarantee that the test attacks are genuinely unseen. In practice, we are left trusting our best guess.

This is what we call \emph{label pollution}: one synthesis system can appear under several obfuscated names across sources (for example \texttt{XTTS}, \texttt{system\_2}, and \texttt{A12}), so a split meant to separate seen from unseen attacks may silently place the same system on both sides. It undermines the trustworthiness of any result reported on combined data, including our own.

Our internal evaluation provides a concrete example of this ambiguity. \autoref{tab:generalisation} reports retrospective results from two older \company detector versions on attacks labeled as seen and unseen. The apparent result is encouraging: accuracy on the nominally unseen attacks matches or exceeds accuracy on the seen attacks. Inspection of the split, however, showed that most of the unseen set came from the \texttt{dev} portions of the training corpora. These samples were new recordings of attack types the detectors had already encountered, not genuinely new attack types. We include the table as an illustrative company case study of how an evaluation split can overstate generalization, not as evidence that the detectors generalize to unseen attacks.

We caught this only because we knew how the set was built. No tooling warned us, and no shared standard says what a valid unseen-attack test must contain. Refining the detector alone cannot establish generalization when the provenance and identity of test attacks are uncertain. Addressing this problem requires both technical methods for tracing or clustering attack sources and shared conventions for testing and reporting generalization (CM1, \autoref{sec:cta}).

\begin{table*}[htbp]
\centering
\begin{tabular}{|l|l|l|l|l|}
\hline
\textbf{System} & \textbf{\begin{tabular}[c]{@{}l@{}}Detection accuracy\\ (seen attacks)\end{tabular}} & \textbf{\begin{tabular}[c]{@{}l@{}}Detection accuracy\\ (unseen attacks)\end{tabular}} & \textbf{\begin{tabular}[c]{@{}l@{}}Balanced accuracy\\ (seen attacks)\end{tabular}} & \textbf{\begin{tabular}[c]{@{}l@{}}Balanced accuracy\\ (unseen attacks)\end{tabular}} \\ \hline
\company v2 & 0.9135 & 0.9381 & 0.9383 & 0.9509 \\ \hline
\company v3 & 0.9723 & 0.9760 & 0.9721 & 0.9745 \\ \hline
\end{tabular}
\caption{Illustrative internal results from two older \company detector versions on attacks labeled as seen and unseen. Detection accuracy is the weighted fraction of clips classified correctly, and balanced accuracy is the mean of the per-class accuracies. The underlying data and importance weights are proprietary; sample counts, exact test composition, operating thresholds, and uncertainty estimates are not available for publication. The table is therefore not intended as a reproducible benchmark, an absolute performance claim, or a comparison between detector versions. Its sole purpose is to show how a nominal seen/unseen split can overstate generalization: most of the ``unseen'' set came from the \texttt{dev} portions of the training corpora and contained new recordings of previously encountered attack types.}
\label{tab:generalisation}
\end{table*}

\subsubsection{Sensitivity to channel, codec, and augmentation}
\label{sec:eval-robustness}

If one finding from the project deserves emphasis, it is the sensitivity of deepfake detection to channel, codec, and augmentation conditions. In our tests, these effects were large enough that architecture comparisons alone did not predict deployment behavior. The sensitivity cuts in two directions, and both matter for deployment.

The first is the miss side. In one internal comparison, a detector trained without representative codec augmentation showed a miss-rate increase from roughly 4\% to 60\% after one non-aggressive codec pass. Codec-specific augmentation closed most of this gap, whereas general-purpose augmentation had little effect. The underlying sample counts, detector configuration, operating threshold, codec settings, and uncertainty estimates are not available for publication; we therefore report this result as an illustrative deployment observation rather than a reproducible performance estimate. The result nevertheless shows why detectors should be evaluated after realistic processing rather than only on clean synthetic audio~\cite{delgado2026}.

The same mismatch affected false alarms in our narrowband telephone setting. In one internal evaluation, false-alarm rates increased from roughly 5\% to 70\%, while equal-error rates reached approximately 16\% and 25\% under AMR-NB and G.711 processing. Adding telephony-representative data and codec augmentation substantially improved the system. As above, the detailed sample composition, detector configuration, thresholds, codec settings, and uncertainty estimates are unavailable, so these values are illustrative internal measurements rather than benchmark results.

These opposite failure modes are themselves a problem for customers. Because each vendor, and sometimes each model, responds differently to the same manipulation, a deployment is tuned to a specific system, and vendors are not interchangeable, even though the technology requires frequent updates. How to keep behavior stable, particularly decision thresholds, across those updates and changing channel conditions, is rarely discussed in the industry.

Provenance matters as much as the model, on both sides of the task. For a fixed attack architecture, detectability varied with the data the attack system itself had been trained on, for example, across its language-specific variants; attack systems that drew on more unified sources, such as Common Voice~\cite{ardila2020commonvoice} or FLEURS~\cite{conneau2023fleurs}, did not show this inconsistency. Taken together with the augmentation and channel findings, deepfake detection is especially fragile to mismatches along three axes: between the detector's own training and test data, in the training data of the spoof systems it must detect, and across the channels of both bona fide and spoofed audio. These observations show that deployment robustness depends on both model design and the match among detector data, spoof-system provenance, and channel conditions.





\subsubsection{Calibration and operating points}
\label{sec:calibration}

Calibration ties all of this to a decision, and it is where the absence of customer data hurts most. The advice we received was unanimous: a deployable detector must be calibrated, ideally recalibrated for the specific use case. The catch is that calibration needs labeled data from the deployment, but customers rarely have it. As a concrete example, one outcome of the project was a detector deployed for \useragency{}, which, like most customers, has no operational labeled data we could calibrate against. The resolution was unsatisfying but unavoidable: we ship a default calibration that we have reason to believe is \emph{good enough}.

The same gap affects operating points. A customer usually needs more than the default threshold; they need to adjust it to match their tolerance for misses versus false alarms. Without their data, we can tell them how to move it, but not where to. And the loop is rarely closed from the other side either: the detector is generally handed over to the customer to deploy as they see fit, with no evaluation on their side, so the mismatch between our test conditions and their real ones may never be measured.

\subsection{Score Communication and Interpretability}
\label{sec:explainability}

The detector normally outputs a calibrated log-likelihood ratio (LLR), along with an explanation of its meaning. But what do the customers do with it? In practice, LLR is opaque to a non-expert: customers collapse it to a single rule, ``\textit{above threshold, therefore a deepfake}'', and discard the graded information the score was meant to carry. The field lacks an agreed convention for turning a calibrated score into a statement a customer can act on, and LLRs in particular sit well outside most users' intuition. Human studies reinforce the importance of context and communication: recognition depends on prior warning and perceived speech quality, and listeners may fail to react to a synthetic message when they do not expect one~\cite{malinka2024human}.

This leaves us in a standoff. Customers would prefer a hard decision, a plain yes or no. Vendors are reluctant to supply one, because once the system issues the verdict rather than the evidence, responsibility for that verdict shifts from the customer to the vendor: if the call is wrong, the vendor ``decided for them''. The output LLR is too abstract for the customer, yet the binary decision the customer wants is one that no vendor wants to provide.

Two further complications make a clean answer hard. First, a calibrated score is meaningful only relative to the data the system was calibrated on; it is not a true (log-)probability that a recording is forged, and presenting it as one is misleading. Second, different audiences need different things: a forensic examiner, a fraud analyst, and an end user ask different questions of the same score, and the feature-level explainability methods developed for image and text classifiers~\cite{lundberg2017shap} transfer poorly to a few seconds of audio; although such methods have been adapted to spoofing detection~\cite{ge2022shap}, doing so remains difficult.

One direction we find promising borrows from speaker identification, where a match is always expressed relative to a reference population~\cite{gonzalez2007emulating}. By analogy, a detector could report not ``\textit{this is deepfake}'' but ``\textit{this is deepfake relative to a stated reference}'', whether that reference is all known speech, a single language, or a narrower subset, and attach a probability once the reference set is fixed. This makes the conditional nature of the score explicit and gives the customer something specific to reason about, even though it remains vulnerable to the same human tendency to collapse the result into a verdict\footnote{By this we mean the human tendency to discard the qualifying context and read a conditional score as a definitive yes or no.}. We have not settled on the design and raise it as another open problem in our call to action (CM3, \autoref{sec:cta}).

\subsection{Deployment and Integration}
\label{sec:integration}



We have comparatively little to report here, and the reason is itself a finding. The detector is handed to the customer as a self-contained module, which the customer then deploys, tunes, and operates on its own. Latency, throughput, and deployment topology, whether on-premises or in the cloud, are consequently set by the customer and are largely outside our visibility. These are real constraints on what a deployable detector can be, yet the prevailing delivery model leaves the vendor little ability to ensure the system meets them during real operation.

The one integration question we can speak to is structural. Customers typically buy deepfake detection alongside speaker identification (SID) as a package, yet the detector is a standalone module with no coupling to the SID system. How best to combine the two so that a speaker identity decision accounts for the possibility of spoofed input remains an open problem we have not yet addressed with the new product. Spoofing-aware speaker verification (SASV)~\cite{SASV} is a natural starting point for future work.

\subsection{Customer Acceptance}
\label{sec:acceptance}





Acceptance criteria are usually negotiated in a contract and typically assume the customer can provide representative data for testing and calibration. Most cannot. Customers rarely have the labeled data needed to define meaningful acceptance criteria, so the criteria are worked out jointly rather than specified up front. Expectations also differ on something as basic as what ``accuracy'' should mean.

Customer acceptance testing further underlines the issue. Customers usually try the first input they happen to have, regardless of whether it falls within the scope the detector was built for, and in our experience, they tend to evaluate the system on data quite different from what they will actually encounter in practice. Acceptance, therefore, rests on evidence that says little about how the system will behave once deployed.

A further reason the process stays loose is that many customers do not \textit{yet} have an operational need for the detector. They are not running it under real load; they only want to be prepared for when deepfake attacks become relevant for them.


\section{The Governance Gap}
\label{sec:regulatory}

This section presents an operational reading of the regulatory situation, not a legal determination. Whether a deepfake detector falls within the EU AI Act's high-risk regime depends on its intended purpose, the role it performs in a larger system, and the context in which it is deployed~\cite{aiact}. In our reading, a standalone detector does not map unambiguously to the biometric use cases listed in Annex~III because it classifies an utterance rather than identifying a person. Article~6(3) may also exclude an Annex~III system from the high-risk category when it does not pose a significant risk to health, safety, or fundamental rights and meets the conditions stated in that provision~\cite[Art.~6(3)]{aiact}. These conditions require case-specific assessment. We therefore cannot conclude that detectors used for fraud screening, forensic analysis, or law enforcement generally fall inside or outside the high-risk regime. The Commission's classification guidance remains non-binding, and its examples are not exhaustive~\cite{commission_guidelines_2026}.

The GDPR adds further context-specific obligations. Its data-minimization principle requires personal data to be adequate, relevant, and limited to what is necessary for the stated purpose~\cite[Art.~5(1)(c)]{gdpr}; it does not by itself determine what training-data provenance a detector provider must publish. Applicable duties depend on factors such as the data involved, the legal basis for processing, the controller's role, and national or sector-specific law. Regulation, contractual governance, and audits may therefore improve data and deployment practices, but their applicability and effects vary across use cases. Our narrower conclusion is that external requirements alone do not create the shared datasets, evaluation protocols, or score-communication conventions identified in this report. Community coordination should complement, not replace, legal compliance and regulatory oversight.

\begin{table*}[htbp]
\centering
\resizebox{0.9\linewidth}{!}{
\begin{tabular}{p{0.25\textwidth} p{0.52\textwidth} l}
\toprule
\textbf{Experience (Sec.)} & \textbf{Unresolved community-level question} & \textbf{Actions} \\
\midrule
Data sourcing \& provenance (\ref{sec:data}) & Can a detector's training data be verified and held to clean-sourcing standards when no one can audit a deployed model? & CR2, CO2 \\
\addlinespace
Data composition \& drift (\ref{sec:data}) & Which data does a detector actually need, how diverse must bona fide speech be, and how do we keep pace with drift? & CR1, CO1 \\
\addlinespace
Generalization (\ref{sec:evaluation}) & How do we know an evaluation tests genuinely unseen attacks and conditions, rather than familiar ones? & CM1 \\
\addlinespace
Metrics \& comparability (\ref{sec:evaluation}) & What metrics and reporting make results meaningful at a deployment operating point, and comparable across conditions? & CM2, CO1 \\
\addlinespace
Score communication (\ref{sec:explainability}) & How should a calibrated score become a decision, which a non-expert can act on and defend? & CM3 \\
\addlinespace
Deployment \& defense (\ref{sec:integration}, \ref{sec:acceptance}) & How should the field prepare for unseen attacks and define ``fit for purpose'' for consequential uses when customers lack representative data and deploy systems independently? & CO3, CO4 \\
\bottomrule
\end{tabular}}
\caption{From experience to roadmap. Each project observation (left) motivates a question for broader investigation (centre), which we map to proposed actions in \autoref{sec:cta} (right): CR = research, CM = method to develop and adopt, CO = coordination.}
\label{tab:roadmap}
\end{table*}

\section{From Experience to Roadmap}
\label{sec:roadmap}

We encountered the problems outlined above in a single project and therefore treat them as candidate recurring challenges rather than universal findings. \autoref{tab:roadmap} separates the observations from our project from the broader questions they motivate and maps each question to a proposed action. These proposals require validation across other vendors, users, jurisdictions, and deployment settings.

\subsection{Call to Action}
\label{sec:cta}

Our experience suggests that the obstacles to deployable deepfake detection have both technical and collective components. We group the resulting actions into three kinds: research challenges (\textbf{CR}), which require new methods or knowledge; methods the field must both develop and adopt (\textbf{CM}), where progress is inseparable from agreement; and coordination actions (\textbf{CO}), which require collective action. This roadmap does not treat benchmark accuracy as sufficient; it includes research on data selection and provenance, methods for testing generalization and for communicating scores, and coordination on shared data and evaluation standards. Coordination runs through all three groups because technical progress depends on data and benchmarks that individual organizations cannot create or validate on their own.

\smallskip
\noindent\textbf{Research challenges.} Open problems; no amount of coordination substitutes for solving them.
\begin{itemize}
    \item \textbf{CR1 -- Principled training-data selection.} Given the data that exists, which subset actually maximizes solution quality, and what is missing? The default today is to ingest everything; we would rather weigh data by importance and prune what does not help. A framework that assesses coverage gaps and re-weighs or filters redundant data, and that copes with the fact that sources drift over time, e.g., when a commercial synthesizer or a video platform's codecs change between versions, would all turn dataset assembly from guesswork into engineering.

     \item \textbf{CR2 -- Verifiable training-data provenance.} A deployed model does not reveal which data were used to train it, so external parties cannot readily evaluate data-sourcing claims. We need methods that make such claims testable under realistic conditions, for example through membership inference, watermarking, or provenance attestation. These methods would complement contractual disclosure, auditing, and regulation by providing technical evidence about model provenance.
    
\end{itemize}

\smallskip
\noindent\textbf{Methods to develop and adopt.} Each needs a method that does not yet exist \emph{and} a community willing to adopt it as standard; neither half suffices alone.
\begin{itemize}
    \item \textbf{CM1 -- Unify labeling and make generalization testing verifiable.} Spoof labels differ from one dataset to the next. So when we combine datasets and test across them, we cannot tell whether the test attacks are genuinely new or just familiar ones wearing a different label. Fixing this requires two steps. First, a shared labeling and attack-taxonomy scheme that the whole community uses. Second, methods that reveal which attack classes a dataset actually contains, such as source tracing and attack-family clustering. STOPA provides one concrete starting point by systematically varying generative components and framing source tracing as an open-world attribution task~\cite{firc2025stopa}. Subsequent experiments show that source-tracing performance also depends strongly on how the verification objective is formulated~\cite{firc2026hidden}. For voice clones, that also means tracking the source material and speaker identity they were built from. Until we have this, ``\textit{testing generalization}'' is a claim no one can check.

    \item \textbf{CM2 -- Develop a real-world evaluation and standardize it.} Two problems compound. First, the metrics: EER or DCF weight every sample equally, and each attack family needs its own threshold, so a single figure hides how the system behaves once attacks are mixed. We need metrics built on an explicit attacker model and evaluation under real conditions, including long recordings, codec degradation, partial synthesis, and customer-side noise. Second, comparability: EER and DCF figures land in different places across datasets, so, alone, they tell a customer almost nothing. Biometrics offers a precedent worth borrowing: ISO/IEC~19795 reports error rates across a range of operating points rather than one threshold~\cite{iso19795}. A standardized procedure with a defined minimum set of reported results would make claims comparable and turn vendor datasheets into checkable statements.

    \item \textbf{CM3 -- Make the detector's output intelligible and decision-ready.} A calibrated log-likelihood ratio is meaningless to the people who must act on it; customers reduce it to ``\textit{above threshold, therefore deepfake}''. What they actually need is a defensible operating point, not just the default one, and often a hard decision, which vendors are reluctant to supply because the responsibility for a hard decision then shifts to them. The community should agree on how to communicate scores to non-experts (what an LLR or posterior probability means, when it should not be acted on, how uncertainty is expressed), how to expose adjustable operating points, and how to serve audiences as dissimilar as a trained forensic examiner and a non-expert reviewer. One promising direction, borrowed from speaker identification, is contextual framing (\textit{synthetic relative to which population?}), paired with an explicit probability once the reference set is fixed. Forensic users in particular are beginning to demand this, including segment-level localization of which parts of a recording are synthetic, and the field has no agreed-upon answer.
\end{itemize}

\smallskip
\noindent\textbf{Coordination and governance.} Actions that complement technical research by establishing shared resources, conventions, and responsibilities.
\begin{itemize}
    \item \textbf{CO1 -- Treat bona fide data with the same rigor as spoof data.} Detector research obsesses over the diversity of synthetic attacks while drawing genuine speech from a handful of clean corpora. In our experience, the distribution of bona fide data (its sources, recording conditions, codecs, and compression) significantly affects real-world performance, and a narrow bona fide set silently limits generalization, regardless of how rich the spoof side is. Bona fide diversity and balanced, matched sampling should be first-class dataset design requirements, not afterthoughts.

    \item \textbf{CO2 -- Agree on a standard for what makes a corpus deployable.} A deployable-corpus standard should cover licensing terms, TTS systems, languages, codecs, recording conditions, and explicit bias controls. SCDF illustrates why speaker characteristics belong in this list: detector performance varied across sex, language, age, and synthesizer type~\cite{stanek2025scdf}. These requirements can be specified once at the level of a community standard, rather than negotiated on a paper-by-paper basis. Today, every group builds its own corpus, and the field lacks a shared understanding of what ``fit for commercial use'' actually means.

    \item \textbf{CO3 -- Create a shared, privacy-preserving pool of attack data.} The core difficulty is defending against attackers we have never seen. We call for a common pool, contributed to by large-scale synthesis vendors, security agencies, and incident responders, that uses mechanisms that respect commercial and privacy constraints. These can include sharing feature representations rather than raw audio, federated learning, or homomorphic schemes. An international, independent body that collects, curates, and redistributes incident data would give the field a fighting chance against novel attacks.

    \item \textbf{CO4 -- Define ``\textit{fit for purpose}'' for consequential deployments.} Whether a deepfake detector qualifies as high-risk under the AI Act depends on its intended purpose and deployment context (\autoref{sec:regulatory}). Regardless of its formal classification, consequential uses such as forensic evidence and law-enforcement support require clear expectations regarding evaluation, documentation, human oversight, and disclosure of failure modes. The community should develop shared technical criteria for these uses without attempting to predetermine their legal classification. In the absence of such criteria, providers must define requirements independently, which limits consistency and comparability.
\end{itemize}

\section{Conclusions}
\label{sec:conclusions}

This report draws on a single industry-academia project and captures only part of the deployment landscape. We do not claim that the frequency or severity of the reported barriers generalizes across vendors, users, jurisdictions, or deployment settings. In our project, achieving strong in-domain performance was only one part of producing a deployable system. Important problems remained in robustness to unseen attacks and channel shifts, calibration, data access, evaluation, score communication, and customer-side validation. These problems combine algorithmic, methodological, operational, and governance dimensions.

The central lesson is not that algorithmic research is complete, but that benchmark accuracy alone does not establish deployability. Further progress requires research on robust generalization, calibration, and data selection, together with shared datasets, evaluation protocols, reporting conventions, and clearer score communication. The research questions and joint actions in \autoref{sec:cta} make these dependencies explicit. If deepfake speech detection is to withstand wider deployment, the community must treat shared infrastructure and technical research as complementary parts of the same problem.

\section{Acknowledgements}
This work was supported by the Ministry of the Interior of the Czech Republic, project VB02000060, and was partially supported by the Brno University of Technology internal project FIT-S-26-9011.

\section{Generative AI Use Disclosure}

During the preparation of this work, the authors used Generative AI Models (specifically Google Gemini, ChatGPT, and Grammarly) for language editing and text refinement. The authors reviewed and edited the output as needed and take full responsibility for the publication's content.

\section{Conflict of interest.} \company{} is a commercial vendor of the technology discussed here, and the authors include members of its staff.

\bibliographystyle{IEEEtran}
\bibliography{mybib}

\end{document}